\documentclass[prb,preprint,showpacs]{revtex4}
\usepackage{graphicx}
\usepackage[T1]{fontenc}
\usepackage{amssymb,amsmath}
\begin{document}
\title{Antiferromagnetic coupling and enhanced magnetization in all-ferromagnetic
superlattices}
\author{P. Padhan}
\affiliation{Laboratoire CRISMAT, UMR CNRS 6508, 
             6, Bld. du Mar\'echal Juin, F-14050 Caen, France}
\author{W. Prellier\thanks{%
prellier@ensicaen.fr}} 
\affiliation{Laboratoire CRISMAT, UMR CNRS 6508, 
             6, Bld. du Mar\'echal Juin, F-14050 Caen, France}
\author{R.C.\ Budhani}
\affiliation{Department of Physics, Indian Institute of Technology, Kanpur 208016, India}
\date{\today}

\begin{abstract}
The structural and magnetic properties of a series of superlattices
consisting of two ferromagnetic metals La$_{0.7}$Sr$_{0.3}$MnO$_3$ (LSMO)
and SrRuO$_3$ (SRO) grown on (001) oriented SrTiO$_3$ are studied.
Superlattices with a fixed LSMO layer thickness of 20 unit cells (u.c.) and
varying SRO layer thickness show a sudden drop in magnetization on cooling
through temperature where both LSMO and SRO layers are ferromagnetic. This
behavior suggests an antiferromagnetic coupling between the layers. In
addition, the samples having thinner SRO layers (n 
%TCIMACRO{\TEXTsymbol{<} }
%BeginExpansion
\mbox{$<$}%
%EndExpansion
6) exhibit enhanced saturation magnetization at 10 K. These observations are
attributed to the possible modification in the stereochemistry of the Ru and
Mn ions in the interfacial region.
\end{abstract}
\maketitle
\newpage

Transition metal oxides are used to design artificial magnetic structures
with different bilayer configurations consisting of ferromagnetic (FM) -
antiferromagnetic (AFM), AFM - AFM, and FM - FM thin films. On several
combinations of FM-AFM bilayers the magnetic behavior exhibit horizontal and
vertical hysteresis loop shifts,\cite{1,2,3,4} while some AFM-AFM bilayer
systems show unexpected ferromagnetic behavior.\cite{5,6,7} Ke {\it et. al.}%
\cite{8} have observed exchange bias effects in multilayers consisting of
FM-FM bilayers composed of La$_{0.67}$Sr$_{0.33}$MnO$_3$ and SrRuO$_3$.
While Uozu {\it et. al.}\cite{9} have observed an antiferromagnetic exchange
coupling in the FM trilayer consisting of Sr$_{0.7}$Ca$_{0.3}$RuO$_3$ (SCRO)
and La$_{0.6}$Sr$_{0.4}$MnO$_3$. in order to understand the coupling between
the Ru and Mn ions at the interfaces of the two FM media, we have
synthesized a series of superlattices consisting of FM-FM bilayers of La$%
_{0.67}$Sr$_{0.33}$MnO$_3$ (LSMO) and SrRuO$_3$ (SRO), where the LSMO layer
thickness (d$_{LSMO}$) is fixed (\symbol{126}20 u.c.) and the SRO layer
thickness (d$_{SRO}$) is varied. We note that the temperature dependence of
magnetization of these superlattices strongly varies with the SRO layer
thickness when the d$_{SRO}$ is less than 6 unit cells. In addition, these
superlattices show a sudden drop in magnetization at a temperature where
both LSMO and SRO layers are ferromagnetic. The reproducible
zero-field-cooled (ZFC) magnetic minor hysteresis loop shape in the
field-cooled (FC) state of these superlattices suggests the existence of
antiferromagnetic exchange coupling but not the presence of an interfacial
layer of antiferromagnetic character.

Thin films of LSMO and SRO and their superlattices were grown on (001)
oriented SrTiO$_3$ (STO) substrates at 720${{}^{\circ }}$C in oxygen ambient
of 300 mTorr using a multitarget pulsed laser deposition technique. The
deposition rates (typically \symbol{126}0.26 \AA /pulse and \symbol{126}0.31
\AA /pulse) of SRO and LSMO respectively were calibrated individually for
each laser pulse of energy density \symbol{126}3 J/cm$^2$. The chamber was
filled with oxygen of 300 Torr after completion of deposition and then the
samples were cooled to room temperature at the rate of 15 ${{}^{\circ }}$%
C/min. The superlattice structures were synthesized by repeating 15 times
the bilayer consisting of a 20- unit cells thick LSMO layer and n- unit
cells thick SRO layer, with n taking integer values from 1 to 12. In all
multilayer samples, the bottom and top layers are of LSMO. Characterization
of the structure and epitaxial nature of the multilayer and single layer
films were performed using x-ray diffraction (XRD). For the magnetization
(M) measurements, we have used a superconducting quantum interference device
based magnetometer (Quantum Design MPMS-5). These measurements were carried
out by cooling the sample to the desired temperature in the presence/absence
of a magnetic fields applied along the [100] and [001] directions of the STO
substrate. The orientation of the magnetic field during the field-cooled
measurements remained the same.

The lattice parameter of cubic STO (3.905 \AA ) is smaller than that of the
pseudocubic lattice parameter of SRO (3.93 \AA ) but larger than that of the
LSMO (3.88 \AA ). Thus, STO provides in-plane tensile stress for the
epitaxial growth of LSMO with a lattice mismatch of -0.64 \%. Similarly, it
is expected that the LSMO would provides in-plane compressive stress for the
epitaxial growth of SRO with a lattice mismatch 1.28 \%. However, the
LSMO-SRO superlattice stabilizes pseudocubic phases of these perovskites. In
the conventional $\Theta $-2$\Theta $ scans of the superlattices, no peaks
were observed other than the (00l) Bragg reflections of the constituents,
the substrate and the satellites due to chemical modulation present in the
multilayer. To evaluate the film thickness of LSMO and SRO layers, we have
carried out quantitative refinement of the $\Theta $-2$\Theta $ scan of the
trilayer structures using DIFFaX program.\cite{10} The experimental $\Theta $%
-2$\Theta $ scans and the simulated profiles of the two samples recorded
around the (001) reflection of STO are shown in Fig. 1. The simulated
profile using the calibrated thickness is in good agreement with the
position of the Kiessig fringes and their relative intensity ratio
confirming the quality of the layers.

LSMO exhibits positive spin polarization and ferromagnetic ordering with a
relatively high Curie temperature (T$_C$\symbol{126}360 K), small magnetic
anisotropy and low coercive field (H$_C$).\cite{11} While SRO shows negative
spin polarization with a relatively small T$_C$ (\symbol{126}150 K), strong
uniaxial crystalline anisotropy and relatively large H$_C$.\cite{12} In the
magnetic multilayers discussed here the thickness of the LSMO layer is fixed
at 20 u.c. A typical field-cooled temperature-dependence magnetization M(T)
for \symbol{126} 20 u.c. thick LSMO and SRO films is shown in the Fig. 2a
and 2b, respectively. These relatively thin films of LSMO and SRO show a
reduced T$_C$ of \symbol{126} 330 K and \symbol{126} 145 K respectively. The
reduced T$_C$ seen in both the cases is due to finite size effect resulting
from the strains.\cite{13}

Fig. 3 displays the field-cooled (FC) magnetization of the superlattices
with n=2, 4, 5, 6 and 7. These measurements were performed at 0.01 tesla
field applied along the out-of-plane direction of the STO. From this figure
it appears that the alternative stacking of LSMO and SRO in a superlattice
changes the M(T) significantly compared to the M(T) of the constituents
(e.g. LSMO\ and SRO). Similar behavior of magnetization is obtained when the
magnetic field is applied in the plane (not shown), but the magnetization is
larger in the case of the out-of-plane magnetic field. The onset of
spontaneous magnetization in all superlattices occurs at T=340 K. The
behavior of M(T) at T%
%TCIMACRO{\TEXTsymbol{<}}
%BeginExpansion
\mbox{$<$}%
%EndExpansion
340 K is significantly different from that of the pure LSMO. For example,
the FC magnetization of the sample with n=2 decreases gradually as the
temperature is raised from 10 K to 33 K. This trend is followed by an
increase of magnetization till a maximum value is reached at 100 K. A
further increase in temperature leads to a monotonic drops in magnetization
(see Fig. 3a). As the SRO layer thickness is increased up to 4 u.c., the
magnetization increases slowly above 10 K, becomes maximum at 60 K, and then
above 60 K decreases slowly. Between 160 K and 200 K, the magnetization
shows a plateau. Two characteristic temperatures can be identified in Fig. 3
in addition to the T$_C$ (350 K) above which the sample becomes
paramagnetic. The first one is the temperature (T$_C$*) at which the SRO
layer becomes ferromagnetic since the magnetization rises sharply below this
temperature (150 K). The second one, is the temperature (T$_N$) below which
the magnetization decreases although both the components are in the
ferromagnetic state. To clarify this point we have enlarge the M(T) curve in
this region and shown them in Fig.3, panel f, g, h, i, and j for
superlattice with n=2, 4, 5, 6 and 7, respectively.

The distinct cusp in magnetization below the T$_C$ of SRO is an indication
of an antiferromagnetic behavior. Thus, we denote the temperature associated
to this feature as the N\'{e}el temperatures (T$_N$). However, when the SRO
layer thickness is increased beyond 7 u.c., this cusp-like feature below the
Curie temperature of SRO is suppressed. The T$_C$ (LSMO), T$_C$* and T$_N$
of the superlattices extracted from Fig. 2 are plotted as a function of d$%
_{SRO}$ in Fig. 4. As discussed before, the T$_C$ of the superlattices is
nearly independent of d$_{SRO}$, and is close to the T$_C$ of LSMO (\symbol{%
126}340K), while the T$_C$* increases with d$_{SRO}$.\ However, this T$_C$*
is not distinguishable for the superlattices with d$_{SRO}$%
%TCIMACRO{\TEXTsymbol{>}}
%BeginExpansion
\mbox{$>$}%
%EndExpansion
8 u.c. On the contrary, for the samples with n%
%TCIMACRO{\TEXTsymbol{<}}
%BeginExpansion
\mbox{$<$}%
%EndExpansion
5 u.c., the T$_N$ shows a distinct increase with the SRO layer thickness.
The presence of a distinct T$_C$(SRO) at 150 K in the M(T) data of the
superlattice with $n\geqslant 4$ indicates the formation of a stoichiometric
SRO layer. This observation suggests that in this magnetic system the
interface roughness caused by the magnetic and structural disorder is small (%
\symbol{126}2 u.c.). The reproducible ZFC in-plane and out-of-plane minor
hysteresis loops of these samples in their corresponding FC state indicate
the existence of an antiferromagnetic exchange coupling, but do not show the
presence of an interfacial antiferromagnetic layer.\cite{8} The drop in
magnetization, at a temperature which we have marked T$_N$, could be due to
this disordered interface. A similar AFM exchange coupling has been observed
in LSMO/SCRO/LSMO trilayers by Uozu et. al.\cite{9}, and in LSMO/SRO
bilayers by Ke and coworkers \cite{8}. While the former group has attributed
the AF coupling to superexchange interaction, Ke et. al. attribute it to
interfacial charge transfer. However, the AFM exchange coupling with the SRO
layer thickness as seen in the present case suggests the possibility of
other physical processes. Some well established sources affecting the
magnetic coupling in superlattices are the substrate-induced strains,
interfacial stress and interlayer exchange coupling.\cite{13,14} The larger
value of T$_C$* compare to the transition temperature of SRO could also be
due to the high paramagnetic susceptibility\cite{15} of SRO above the T$_C$%
(SRO) and/or the reconstruction of the spin state of Ru and Mn ions at/close
to the interfaces. The value of the T$_N$ increases and saturates to a
temperature T$_C$(SRO) at higher d$_{SRO}$, indicating the influence of the
size effect of the SRO layer.

In order to study the possible magnetic configuration of Ru and Mn ions at
10 K, we have measured the ZFC magnetic hysteresis loop of these samples
with the magnetic field oriented along the [100] and [001] directions of the
substrate. The in-plane and out-of-plane ZFC magnetic hysteresis loops of
two samples are shown in Fig. 5a. The in-plane and out-of-plane saturation
magnetizations (M$_S$) of all samples are the same though their in-plane and
out-of-plane saturation magnetic fields are different. The M$_S$ of some
samples, extracted from their ZFC in-plane and out-of-plane hysteresis loop
after correcting for the weak diamagnetic response of the substrate, are
shown in the Fig. 5b. This figure also includes the theoretical value of M$%
_S $ calculated from the spin-only M$_S$ of the LSMO (3.34 $\mu _B$/Mn)\cite
{16} and SRO (1.6 $\mu _B$/Ru)\cite{17}. The higher value of the measured M$%
_S$, as compared to the theoretical one for the sample with $n=2$ to 7,
indicates an enhancement of the magnetization.

The enhance magnetization could be due to the modification of the charge
states of the Ru and Mn ions\cite{18} at the interfaces, and thereby an
increase in the effective thickness of the interfacial layer in superlattice
with the lower d$_{SRO}$ ($<$ 4 u.c.). In samples with d$_{SRO}$ $>4$, the
effective thickness of the interfacial layer decreases as the stoichiometric
SRO layer start to form. The interfacial magnetic roughness seems to
suppressed completely by the strong long range ordering of the ferromagnetic
moments of SRO and LSMO at still larger d$_{SRO}$. However, measurements
using electron energy loss spectroscopy should be perform to verify the
valency assumptions, in particular the stabilization of the Ru$^{5+}$ spin
state.

We have demonstrated that in La$_{0.7}$Sr$_{0.3}$MnO$_3$ and SrRuO$_3$
superlattices an antiferromagnetic coupling and increase magnetization can
be induced by changing the SRO layer thickness. We attribute these changes
to interfacial magnetic and electric disorder, which appears to heal as the
SRO layer thickness increases and long range ordering of the magnetic moment
associated with Ru ions becomes dominant.

We thank the financial support from the Centre Franco-Indien pour la
Promotion de la Recherche Avancee/Indo-French Centre for the Promotion of
Advance Research (CEFIPRA/IFCPAR) under Project No 2808-1. Partial support
from the European Union under the STREP\ research project CoMePhS (N${%
{}^{\circ }}517039$) is also acknowledged.

\newpage

Figure captions:

Figure 1: The experimental (solid line) and simulated (DIFFaX) (dotted line)
- 2 x-ray diffraction profiles of the superlattice with n=4 and 8. The (001)
Bragg's reflection of STO and several orders $(0,\pm 1,\pm 2)$ of satellite
peaks are indexed.

Figure 2: Temperature dependence of the out-of-plane magnetization under
0.01 tesla field for a 20 u.c. thick film of LSMO (panel a) and SRO (panel
b).

Figure 3: Temperature dependence of the out-of-plane magnetization under
0.01 tesla field of the superlattices with n=2, 4, 5, 6 and 7 (panels a, b,
c, d and e, respectively, for 0-300K\ and panels f, g, h, i, and j,
respectively, for the enlarge part corresponding to the temperature range
where the LSMO and SRO are ferromagnetic (panel f, g, h, i, and j
respectively). The arrows indicate the T$_C$ and T$_N$.

Figure 4: Evolution of the T$_C$, T$_C$* and T$_N$ (panel a, b and c
respectively) of several superlattices as a function of the SRO layer
thickness. In panel a and b the solid lines represent the T$_C$ of bulk LSMO
and SRO, respectively, while in panel c the solid line is only a guide to
the eyes. Both T$_C$ and T$_C$* have been calculated from the intersection
of the slope around the transition of magnetization.

Figure 5: (a) ZFC magnetization loop for the (20 u.c.)LSMO/(n u.c.)SRO
superlattice with n = 4 and 12. Magnetic field is oriented along the [100]
and [001] directions of STO. (b) : Experimental and theoretical saturation
magnetization for the superlattices as a function of the SrRuO$_3$ layer
thickness. The solid lines are guide to the eyes.

\end{document}